\documentclass[a4paper,11pt]{article}
\usepackage[utf8]{inputenc}
\usepackage[T1]{fontenc}
\usepackage{lmodern}
\usepackage{geometry}
\usepackage{graphicx}
\usepackage{authblk}
\usepackage{url}
\usepackage{cite}
\usepackage{fancyhdr}
\usepackage{amsmath}
\usepackage{braket}
\usepackage{subcaption}

\usepackage[hidelinks]{hyperref}
\usepackage{orcidlink}

\fancypagestyle{firstpage}{
	\fancyhf{}
	\fancyhead[C]{\includegraphics[height=2.2cm]{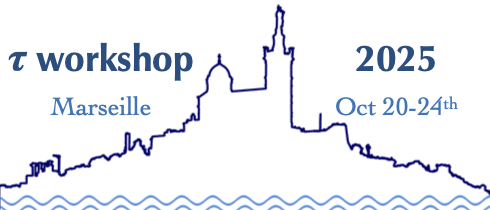}\\ \textit{Proceedings of the 18th International Workshop on Tau Lepton Physics}}

}

\title{Beyond Form Factors: Precise Angular
Tests in Hadronic $\tau$ Decays}
\author[1]{E. Estrada\,\orcidlink{0009-0007-3360-7908}} \author[2,3]{E. Passemar\,\orcidlink{0000-0002-8553-6159}}
\author[2]{S. Paz\,\orcidlink{0009-0005-8763-7171}\thanks{Presenting author.}} \author[4]{A. Rodríguez-Sánchez\,\orcidlink{0000-0001-7291-2146}} \author[1]{P. Roig\,\orcidlink{0000-0002-6612-7157}}

\affil[1]{Departamento de Física, Centro de Investigación y de Estudios Avanzados del
Instituto Politécnico Nacional
Apartado Postal 14-740, 07360 Ciudad de México, México}
\affil[2]{Departament de Física Teòrica, Instituto de Física Corpuscular, Universitat de València — Consejo Superior de Investigaciones Científicas,
Parc Científic, Catedrático José Beltrán 2, E-46980 Paterna, Valencia, Spain}
\affil[3]{Physics Department, Indiana University, Bloomington, IN 47405, U.S.A.}
\affil[4]{Departamento de Física, Universidad de Castilla-La Mancha,
Avenida de Carlos III, s/n, 45004 Toledo, Spain}

\date{}

\begin{document}

\maketitle
\thispagestyle{firstpage}

\begin{abstract}
	Semileptonic $\tau$ decays mainly proceed via interactions between charged lepton and quark currents. The hadronization of the quark current is intrinsically nonperturbative and generally cannot be addressed analytically. In these proceedings, we propose using symmetry arguments alone to construct clean angular observables, which, within the Standard Model and in the absence of long-distance electromagnetic corrections, remain form-factor independent. These predictions can be experimentally tested, and any observed deviation could signal either effects of physics beyond the Standard Model or provide a clean benchmark for long-distance electromagnetic corrections. We also perform a first estimate of the expected impact of new physics in an EFT framework.
\end{abstract}

\section{Introduction}

The $\tau$ lepton, being the heaviest and the only lepton capable of decaying into light pseudoscalar mesons, is a perfect laboratory for the study of weak interactions and, in general, for probing the Standard Model (SM)~\cite{Pich:2013lsa}. Hadronic $\tau$ decays are particularly interesting since they allow us to analyze strong interactions and hadronization in the final state with a simpler initial state and relatively cleaner observables than purely hadronic processes~\cite{Rodriguez-Sanchez:2025nsm}. Traditionally, the preferred approach to describing the hadronization of the quark current is to parametrize it in terms of so-called form factors, which we do not know how to predict from first principles in general. While meaningful information about them can be obtained from nonperturbative approaches such as chiral expansions, dispersion relations, or constraints from lattice QCD data, systematic uncertainties are typically hard to estimate.

In contrast with previous analyses, aimed at better understanding the hadronic dynamics involved in the form factors, we propose to analyze what can be predicted independently of them, thus circumventing, at least to a certain extent, the need to estimate those uncertainties. This can be achieved, starting from general angular distributions, by finding relations in the decay spectra that remain mostly form-factor independent. There are several motivations to pay special attention to this type of observables. First, being able to make first-principles predictions involving the interplay of different fundamental forces that can be experimentally tested is interesting by itself. Second, experimental systematic uncertainties involving hadrons are also hard to control, and underestimated effects on this side have led to some anomalies in the past, such as discrepancies in the muon anomalous magnetic moment $a_\mu$ or in tests of lepton universality. These theoretically clean observables can be useful benchmarks to check that those uncertainties are under control. On the other hand, the results are theoretically clean only up to a certain extent, typically at the percent level, where the interplay with electromagnetic corrections enters into the relations. Understanding these effects is a topic of ongoing research, and this type of observables can provide a clear laboratory~\cite{Cirigliano:2002pv, Escribano:2023seb, Antonelli:2013usa}. Finally, we also consider potential new physics (NP). A useful framework for testing the SM at low energies is the use of effective field theories (EFTs). They incorporate dimension-$D>4$ non-standard interactions (NSIs) generated by heavy beyond-the-Standard-Model (BSM) mediators, without specifying the complete UV theory~\cite{Cirigliano:2009wk}. In particular, the Weak EFT (WEFT) we use also integrates out SM heavy particles such as the $W$ and $Z$ bosons, as well as the top quark. Within this framework, we illustrate how some of the relations are genuine SM predictions, and how they may be modified in different BSM scenarios.

Experimental distributions and branching fraction determinations for $\tau$ decays have been largely studied in experiments like Aleph~\cite{ALEPH:1990ndp}, Belle~\cite{Belle:2000cnh} and BaBar\cite{BaBar:2001yhh}. Below we will make use information from the provided spectra without making further assumptions on the form factors. This can be further extended to a polarized analysis that might be interesting in future experiments like the proposed polarization upgrade in chiral Belle~\cite{Roney:2025lwo}.

\section{WEFT and form factors}

We will parameterize NP using the WEFT. Specifically, the part of the lagrangian that corresponds to $\tau$ decays into light quarks $u$ and $D=d,s$, without lepton flavour or number violation is shown in Eq.~\eqref{eq:lag}.
\begin{align}
\label{eq:lag}
    \nonumber\mathcal{L}_{eff}^{uD}&= -2\sqrt{2}G_\mu V_{uD}\Big[\left(1+\epsilon_L^D\right)\left(\overline{\tau}\gamma_\mu P_L\nu_\tau\right)\left(\overline{u}\gamma^\mu P_LD\right)+\epsilon_R^D\left(\overline{\tau}\gamma_\mu P_L\nu_\tau\right)\left(\overline{u}\gamma^\mu P_RD\right)\\&+\frac{1}{2}\left(\overline{\tau}P_L\nu_\tau\right)\left(\overline{u}\left(\epsilon_S^D-\epsilon_P^D\gamma^5\right)D\right)+\frac{1}{4}\hat{\epsilon}_T^D\left(\overline{\tau}\sigma_{\mu\nu}P_L\nu_\tau\right)\left(\overline{u}\sigma^{\mu\nu}P_LD\right)\Big]+h.c.
\end{align}

In this lagrangian density, we have introduced 5 new couplings on top of the SM effective coupling from the Fermi four-fermion interaction. These interactions are put in terms of the Fermi constant $G_\mu$ measured in muon decays. As a proof of concept, we are restricting this present work to $\tau$ decays into two pseudoscalars. This means that the only relevant parameters will be $\epsilon_S^D,~\hat{\epsilon}_T^D$ and the combination $(1+\epsilon_L^D+\epsilon_R^D)$ that serves as a rescaling of the couplings. The axial and axial-vector components of the current do not contribute to the amplitude due to parity. The scalar and tensor currents introduce additional form factors that have to be introduced with their corresponding normalization factor and Lorentz structure. Additionally, the former form factor can be related to the longitudinal component of the vector current using the equations of motion. They are described in Eq.~\eqref{eq:ff's}, following the convention in Ref.~\cite{Aguilar:2024ybr}, where $p_-$ and $p_0$ correspond to the charged and neutral pseudoscalar momenta, respectively, $q=p_-+p_0$ and $\Delta_{PP'}=m_{P}^2-m_{P'}^2$. 

\begin{align}
\label{eq:ff's}
    \nonumber H &= \bra{PP'}\overline{D}u\ket{0} = F_S^{PP'}(s) = C_{PP'}\frac{\Delta_{PP'}}{m_D-m_u}F_0^{PP'}(s)~, 
    \\
    \nonumber H^\mu&=\bra{PP'}\overline{D}\gamma^\mu u\ket{0} = C_{PP'}\left\{p_{T}^\mu F_V^{PP'}(s)+q^\mu\frac{\Delta_{PP'}}{s}F_0^{PP'}(s)\right\}~,
    \\
    H^{\mu\nu} &= \bra{PP'}\overline{D}\sigma^{\mu\nu}u\ket{0} = -iF_T^{PP'}(s)\left(p_{-}^{\mu} p_{0}^{\nu}-p_{-}^{\nu} p_{0}^{\mu}\right)~.
\end{align}

Using a dispersive analysis, we can make an approximation where, given that the phases for $F_T^{PP'}(s)$ and $F_V^{PP'}(s)$ both can be described using the $S=1$ and $I=1$ scattering phase $\delta_1^1$, we can write $F_T^{PP'}(s)=F_T^{PP'}(0)F_V^{PP'}(s)$ up to small corrections in the inelastic region~\cite{Cirigliano:2017tqn}. We will also use the fact that the scalar form factor is heavily suppressed due to the conservation of G-parity in the $\pi\pi$ case. In the $\pi^-K_S$ channel, there is also a suppression of about two orders of magnitude of the scalar contribution~\cite{Bernard:2011ae} to the total decay width, this allows us to perform a similar approximation when constructing the observables, up to a scalar correction. 

We then arrive at the expression for the doubly differential decay width for the process at first order in the couplings (see Ref.~\cite{Aguilar:2024ybr} for full expression). In this result, we have taken the isospin limit, neglecting terms proportional to $\Delta_{PP'}$
\begin{align}
    \nonumber\frac{\mathrm{d}^2\Gamma}{\mathrm{d}s\mathrm{d}\cos{\theta}}&=\frac{m_\tau^3}{512\pi^3}G_\mu^2|V_{ud}|^2C_{PP'}^2\lambda_{PP'}^{1/2}\left(1-\frac{s}{m_\tau^2}\right)^2\left\{(1+\delta_V)\lambda_{PP'}\left[\frac{s}{m_\tau^2}+\left(1-\frac{s}{m_\tau^2}\right)\cos^2{\theta}\right]\left|F_V^{PP'}\right|^2\right. \\
    &\left.+\frac{\lambda_{PP'}s}{m_\tau C_{PP'}}\operatorname{Re}\left(\hat{\epsilon}_T^DF_V^{PP'}F_T^{PP'*}\right)-2\frac{\lambda_{PP'}^{1/2}\Delta_{PP'}}{m_\tau(m_D-m_u)}\operatorname{Re}\left(\epsilon_S^DF_V^{PP'}F_0^{PP'*}\right)\cos{\theta}\right\}~,
\end{align}
\noindent where $\lambda_{PP'}=(1/s^2)(m_P^4+m_{P'}^4+s^2-2m_P^2s^2-2m_{P'}^2s^2-2m_P^2m_{P'}^2)$ is the normalized Källén function,  $1+\delta_V=1+\epsilon_L^D+\epsilon_R^D$ is the vector rescaling, and $\theta$ is the angle between the $\tau$ and the charged pseudoscalar.

\section{Angular observables}

The resulting decay width for the process $\tau\rightarrow PP'$, under these considerations, is a second order polynomial on $\cos{\theta}$. We can then study the even angular moments (since the linear part on $\cos{\theta}$ is precisely the suppressed term in $\pi\pi$ and $K\pi$) defined as 
\begin{equation}
    I_{2n}\equiv\langle \cos^{2n}{\theta}\rangle=\int_{-1}^1\mathrm{d}\cos{\theta}\,\frac{\mathrm{d}^2\Gamma}{\mathrm{d}s\mathrm{d}\cos{\theta}}\cos^{2n}{\theta}~.
\end{equation}

With these moments defined, the $I_0$ being just the measured spectra, we can make our prediction for the second moment in terms of the spectrum
\begin{equation}
    I_2=\frac{I_0}{5}\left(\frac{3m_\tau^2+2s}{m_\tau^2+2s}\right)\left\{1-4\operatorname{Re}\left(\hat{\epsilon}_T^d\right)\frac{F_T^{PP'}(0)m_\tau s(m_\tau^2-s)}{C_{PP'}(m_\tau^2+2s)(3m_\tau^2+2s)}+\mathcal{O}(\hat{\epsilon}_T^2)\right\}~,
\end{equation}
\noindent so that the only contribution from NP is through the tensor coupling. This means that we can isolate the non-standard term and the coupling can be estimated with
\begin{equation}
    \operatorname{Re}\left(\hat{\epsilon}_T^d\right)\propto\left(2+3\frac{m_\tau^2}{s}\right)I_0-\left(10+5\frac{m_\tau^2}{s}\right)I_2~.
\end{equation}
\noindent With a measurement of $I_2$ we could expect a non-zero value for this subtraction to be a signal of NP or of the long-distance electromagnetic corrections as a function of $s$. 

We could also look at the integrated observables for $I_2$ over the allowed region of $s$. Two integrations have been explored: the unweighted integration
\begin{equation}
    J_2\equiv\int_{(m_P+m_{P'})^2}^{m_{\tau}^2}\mathrm{d}s\,I_{2}(s)~=\left(1+a\operatorname{Re}\left(\hat{\epsilon}_T^d\right)\right)\int_{(m_P+m_{P'})^2}^{m_{\tau}^2}\mathrm{d}s\,\frac{1}{5}\left(\frac{3m_\tau^2+2s}{m_\tau^2+2s}\right)I_{0}(s)
\end{equation}
\noindent and the weighted one that has an easier interpretation as a deviation from the expected branching fraction
\begin{equation}
    J_0\equiv\int_{(m_P+m_{P'})^2}^{m_{\tau}^2}\mathrm{d}s\,5\frac{m_\tau^2+2s}{3m_\tau^2+2s}I_{2}=\left(1+b\operatorname{Re}\left(\hat{\epsilon}_T^d\right)\right)\Gamma_{PP'}~.
\end{equation}

\begin{figure}[h]
    \centering
    \begin{subfigure}{0.49\textwidth}
        \includegraphics[width=0.9\linewidth]{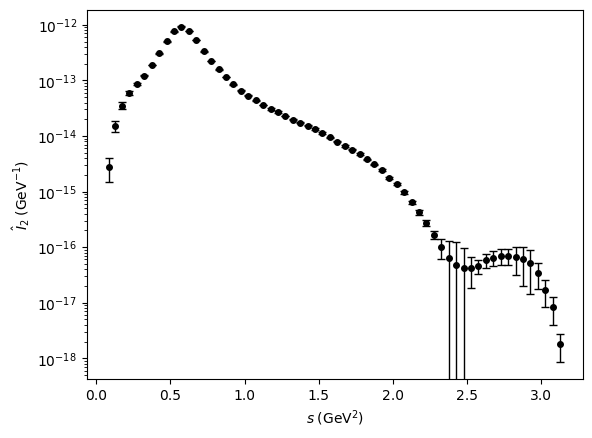}
        \caption{}
        \label{fig_i2a}
    \end{subfigure}
    \hfill
    \begin{subfigure}{0.49\textwidth}
        \includegraphics[width=0.9\linewidth]{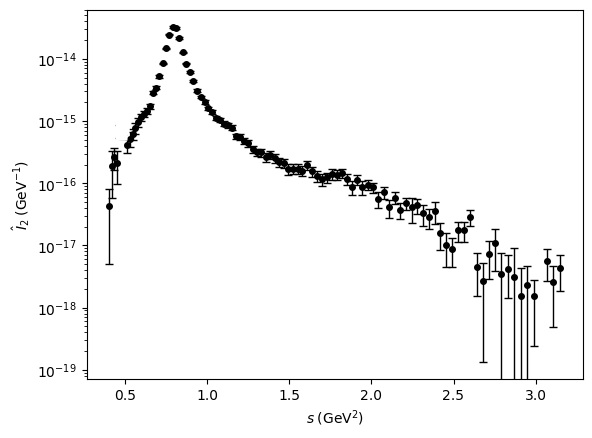}
        \caption{}
        \label{fig_i2b}
    \end{subfigure}
    \caption{SM prediction for the second moment spectra in the $\tau^-\rightarrow\pi^-\pi^0$~(\subref{fig_i2a}) and \\ $\tau^-\rightarrow\pi^-K_S$~(\subref{fig_i2b}) channels using data from Refs.~\cite{Belle_pipi,Belle_kpi}.}
    \label{fig_i2}
\end{figure}

We have found no significant difference in the sensitivity to NP between both integration methods. This sensitivity is encoded in the parameters $a$ and $b$. As an example, the values for the $\pi\pi$ channel are $a=0.300\pm0.011\pm0.030$ and $b=0.301\pm0.011\pm0.030$. The declared uncertainty includes an estimated $10\%$ theoretical uncertainty due mainly to the assumed proportionality between $F_T^{PP'}(s)$ and $F_T^{PP'}(s)$. This estimation was carried out for the $\pi\pi$ and $K\pi$ channels using Belle data for $I_0$ (see Refs.~\cite{Belle_pipi,Belle_kpi}). We treated the spectra numerically as 
\begin{equation}
    I_0^{exp}=\left(\frac{\mathrm{d}\Gamma}{\mathrm{d}s}\right)^{exp}=\frac{\mathcal{B}_{PP'}^{exp}}{\tau_\tau^{exp}}\left(\frac{1}{N_{PP'}}\frac{\mathrm{d}N_{PP'}}{\mathrm{d}s}\right)^{exp}~.
\end{equation}
\noindent The resulting prediction for the second angular moment can be seen in Fig.~\ref{fig_i2} for the $\pi\pi$ and the $K\pi$ channels, after neglecting the scalar contribution. The error bars correspond to the combined statistical and systematic errors in the measurement of the spectra.

\section{Conclusion}

Angular observables are an excellent way to extract information from decay spectra and can be used in hadronic $\tau$ decays to make measurable SM predictions or, if the precision is high enough, to try to isolate radiative corrections. Furthermore, a model-independent analysis such as the one performed through EFTs, along with a form-factor-free hadronic study, is an adequate method for testing the SM and searching for NP.

We have presented a couple of observables as a proof of concept for two channels that benefit from a near isospin-conserving limit, but this work can be extended to further asymmetries, to a polarized analysis, and to additional channels. Given the current precision of the spectra, it may be preferable to study integrated observables rather than the differential width. This highlights the predictive power of angular observables in semileptonic decays, although further theoretical and experimental work is needed to fully exploit it.

\section{Acknowledgments}

This work was supported in part by the FPU scholarship given to SP by the Spanish Ministry of Science, Innovation and Universities under grant FPU24/01729. EE is grateful to Conahcyt/Secithi for funding during his Ph. D. P. R. acknowledged Conahcyt/Secithi funding through the project CBF2023-2024-3226. ARS and EP were supported  by the Generalitat Valenciana (Spain) through the plan GenT programs
(CIDEIG/2023/12 and CIDEGENT/2021/037). This work has also been supported by 
the Spanish Government (Agencia Estatal de Investigaci\'on MCIN/AEI/10.13039/501100011033) Grant No. PID2023-146220NB-I00

\bibliographystyle{JHEP}
\bibliography{references}

\end{document}